# Spatiotemporal Electron Microscopy of Phonon Polaritons in α-$MoO_3$

**Harel Nahari[1,2], Yaniv Kurman[1,2], Raphael Dahan[1,2], Yuval Adiv[1,2], Michael Yannai[1,2], Hanan Herzig Shenfux[3], Frank H.L. Koppens[4,5], and Ido Kaminer[1,2*]**

[1]*Department of Electrical and Computer Engineering, Technion – Israel Institute of Technology, Haifa 32000, Israel*
[2] *The Helen Diller Quantum Center, and The Solid-State Institute, Technion – Israel Institute of Technology, Haifa 32000, Israel*
[3]*Department of Physics, Bar Ilan University, Ramat-Gan, Israel*
[4]*ICFO-Institut de Ciències Fotòniques, The Barcelona Institute of Science and Technology, Castelldefels (Barcelona) 08860, Spain*
[5]*ICREA-Institució Catalana de Recerca i Estudis Avanats, Barcelona 08010, Spain*

***Correspondence**: kaminer@technion.ac.il

**Abstract**: Photon-induced near-field electron microscopy (PINEM) has emerged as a powerful technique for imaging optical excitations with nanometer spatial and sub-picosecond temporal resolution. Recent years have extended the bandwidth of operation of PINEM experiments from the visible range to the mid-infrared, revealing the spatiotemporal dynamics of polaritons and their exotic phenomena. In this study, we nearly double the bandwidth of PINEM, going deeper into the infrared up to 12 µm. Leveraging this advancement, we investigate the spatiotemporal dynamics of phonon polaritons (PhPs) in α-$MoO_3$, a material of growing interest thanks to its in-plane anisotropy. Visualizing PhPs in a cavity-like flake reveals their spatial distribution, dynamics, and wavelength-dependent lifetime. Our work pushes the frontiers of PINEM imaging and highlights its potential for probing hard-to-access polaritonic properties of novel van der Waals materials.

## Introduction

Electron microscopy and spectroscopy are powerful methods for probing optical excitations with nanometer spatial resolution[1,2]. The rapidly evolving field of photon-induced near-field electron microscopy (PINEM)[3,4,5] can enhance the imaging signal by stimulating optical excitations. In the case of a thin sample, electrons can penetrate through it and hence PINEM can access both the buried near-field inside the sample and the external field extending out of it[6,7]. In this context, PINEM efforts extended from imaging of plasmons[3,6] and optical modes of photonic cavities[8,9] to silicon photonic devices[10,11,12,13]. More recent works reveal additional field properties: Lorentz-PINEM, for example, allows reconstruction of the field's phase[14], whereas consequent interactions with the investigated sample and a reference sample enables imaging the field's attosecond oscillations[15] of sub-cycle dynamics[16,17,18].

A growing body of literature in recent years focuses on exploring two-dimensional (2D) materials and understanding their optical excitations[19,20,21], such as phonon-polaritons (PhPs)[22]. PhPs are hybrid light–matter quasiparticles formed by the strong coupling between photons and optical phonons, enabling deeply subwavelength propagation of electromagnetic energy in polar crystals. In this area, studies adapted the PINEM technique to the mid-IR range, reaching wavelengths up to 7 µm. This advancement has enabled recording the spatiotemporal dynamics of PhPs wavepackets[23], as well as the creation and annihilation of PhPs vortices[24] in hexagonal boron nitride (hBN). More recent works used hBN under mid-IR PINEM as a platform to demonstrate coherent amplification[16], allowing the reconstruction of record-low field intensities, and the observation of superluminal dynamics of phase singularities[18].

The burgeoning discovery of new 2D materials in recent years calls for further advances in PINEM technology that would enable probing their exotic photonic properties. In particular, α-phase molybdenum trioxide (α-$MoO_3$) has emerged as one of the more exotic 2D materials[25,26] thanks to its distinctive in-plane hyperbolic nature. This property underlies a range of unusual polaritonic phenomena including reversed Cherenkov radiation[27] and diffraction-less propagation in multilayer structures[28,29], making it a promising platform for photonic applications in the infrared[30,31]. PhPs in α-$MoO_3$ have so far been investigated using s-SNOM[25,26] revealing many of their intriguing properties.

However, SNOM cannot directly access buried near-fields nor the fields extending outside the confines of the sample – features that are intrinsic to PINEM.

In this work, we push forward the capabilities of PINEM, nearly doubling its spectral range to 12 µm. Using the extended spectral range of PINEM, we directly image the PhP mode in an α-$MoO_3$ cavity-like flake (Fig. 1), record its evolution (Fig. 2), and extract its lifetime (Fig. 3). This approach provides unique access to the fields extending outside the sample, as shown in Fig. 1C. Looking forward, extending the capabilities of PINEM could provide access to a plethora of novel physical phenomena including polaritons in many of the famous 2D materials that have been hard to probe by other techniques.

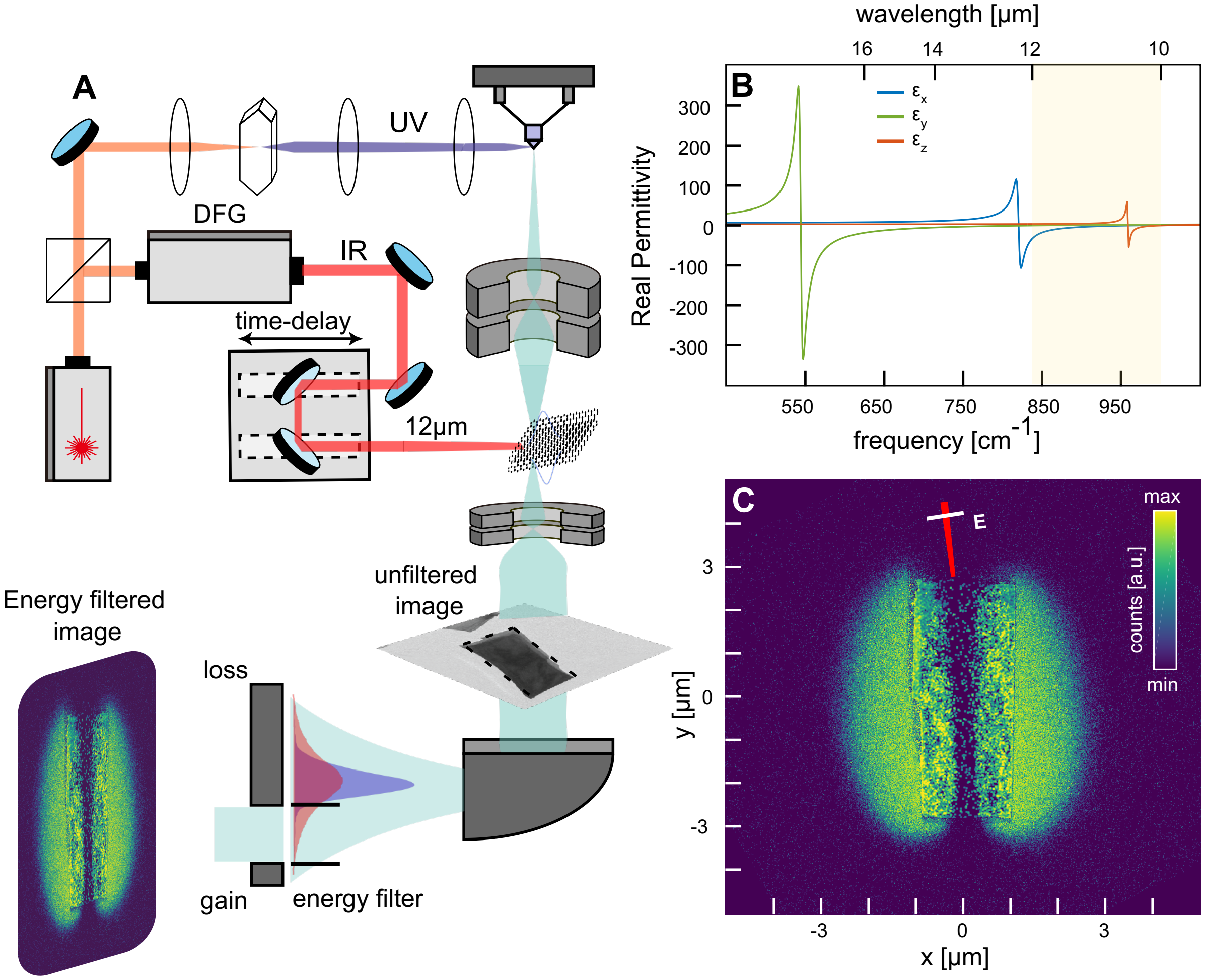

**Fig. 1. Photon-induced near-field electron microscopy (PINEM) of phonon polaritons (PhPs) in α-$MoO_3$. (A)** Experimental setup. A femtosecond laser (orange) splits into two branches. The bottom branch is converted into a mid-IR (red) pulse, tuneable between 10-12 µm, using difference frequency generation (DFG), which then excites the PhP-mode in an α-$MoO_3$ flake. The top branch is converted into an ultraviolet (UV; purple) pulse, around 257 nm, using fourth-harmonic generation, which then photoelectrically excites the electron pulse (cyan). The electron simply images the α-$MoO_3$ sample when counting electrons at all energies (unfiltered image) and images the PhP-mode when counting only electrons that gained energy (filtered image). **(B)** Real part of α-$MoO_3$ permittivity adopted from Ref. 25. In our experiment, the laser excitation wavelength varies between 10-12 µm (833-1000 cm$^{-1}$; opaque yellow) in different measurements. **(C)** The spatial distribution of the PhP-mode in an investigated α-$MoO_3$ flake can be inferred from the filtered electrons. Counts correspond to field strength. The experiment is performed on a 2.5 µm × 7 µm α-$MoO_3$ cavity-like flake with a thickness of 150 nm.

**Results**

At the core of our experiment, probe electron pulses interact with the PhP mode that is excited by pump IR pulses (Fig. 1A). The interaction changes the energy distribution of the electrons, acquiring an image of the PhP mode (Fig. 1C) by energy filtering the electrons. The resolved PhP mode shows its out-of-plane electric field distribution both inside and outside the sample. Fine features of the PhP mode are visible – the nodal line along the center of the flake and the dipole-shape of the lobes outside. Interestingly, this field pattern is strongly constrained by the anisotropic nature of α-$MoO_3$: i.e., the nodal line can appear only along a specific axis (the long axis here, denoted as y), even when we change the illumination angle. This feature is explained by the known hyperbolic nature α-$MoO_3$ ($\varepsilon_y > 0$ and $\varepsilon_x < 0$) for our wavelength range (Fig. 1B), which inhibits all PhP modes with zero $k_x$, and thus maintains a protected nodal line along y.

By changing the time delay between the electron probe and the IR pump pulses, we capture images at different time instances and resolve the spatiotemporal dynamics of the PhP cavity-like mode (Fig. 2A). In the analysis of the field dynamics, we first extract the field profile along the edge of the flake by averaging the signal along the shorter axis (x axis, width of red rectangle at Figure 2A inset) for every time frame. This averaging increases the signal-to-noise ratio and isolates the mode dynamics, which occurs along the long axis (y axis, shown in Fig. 2B). The high sensitivity of PINEM allow us to track the movement of the rather flat peak. Figure 2A denotes the peak location by showing a heatmap of the field intensity across time. Interestingly, the wavepacket initially moves at 2.4 $\mu\mathrm{m/ps}$ (125 times slower than the speed of light in vacuum), and then decelerates when reaching the flake's edge to 0.55 $\mu\mathrm{m/ps}$ (545 times slower than the speed of light in vacuum).

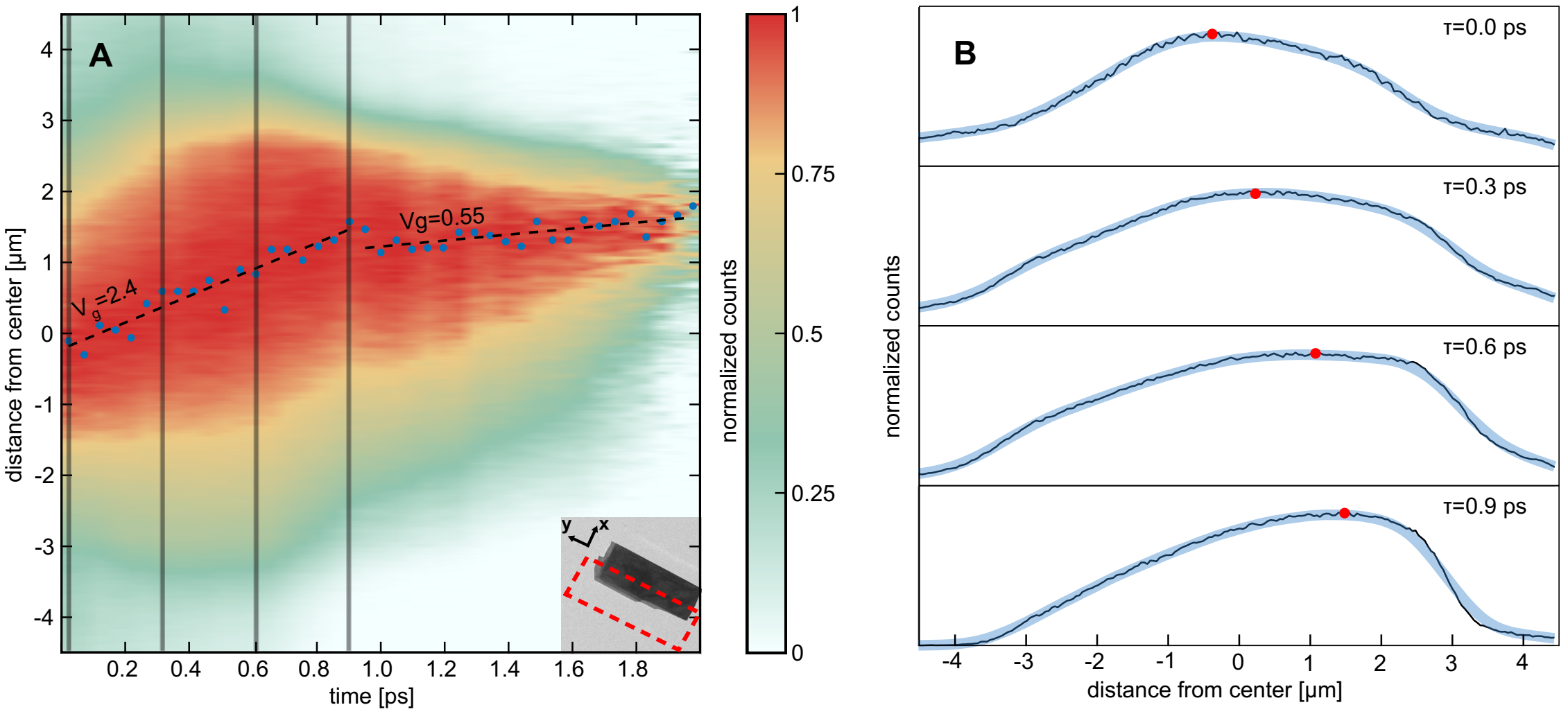

**Fig. 2. Direct observation of the spatiotemporal dynamics of the cavity-like PhP mode in α-$MoO_3$. (A)** Heatmap presenting the measured PhP wavepacket as a function of time and distance from the flake's center. To facilitate tracking of the peak location, the signal is normalized at each point in time to have a peak value of 1. The blue dots represent the peak location, showing clear movement from the center towards the edge of the flake. The group velocities are indicated in units of μm/ps. The inset shows an unfiltered image of the $MoO_3$ flake denoting by the red rectangle the region from which the signal is extracted. **(B)** Measured signal when averaged along the flake's narrow (x) axis, orthogonal to the motion, which is along the long (y) axis, shown for four time frames, corresponding to the vertical lines in panel A. The thicker lines represent the smoothed response; the red points mark the peak of each profile.

Another advantage of PINEM is the ability to perform electron spectroscopy using the same setup. We measure the electron energy spectrum as a function of time delay (representative data shown in Fig. 3A), and calculate the energy variance for each time instance (Fig. 3B). From this data, we can extract the PhP-mode lifetime, by analyzing the electron energy distribution after the interaction:

$$P_k(E,t) = J_k^2\left[2|g|\left(\frac{\Theta(t)e^{-t/\tau}}{\tau}\right) * L(t)\right] * G(t-\zeta E,\sigma_e), \tag{1}$$

where $*$ denotes convolution, $E$ is the electron energy, $t$ is the time delay between the pump and the probe, and $J_k$ is the $k^{\text{th}}$-order Bessel function of the first kind. $L(t)$ is the IR-laser pulse temporal shape, which we measure using a reference sample (a mirror). $G(t-\zeta E,\sigma_e)$ accounts for the electron Gaussian pulse duration of width $\sigma_e$ and chirp $\zeta$, $\Theta(t)$ is the Heaviside step function, and $\tau$ is the PhP-mode lifetime we aim to extract. $g$ represents the PhP-mode amplitude converted into a dimensionless quantity by the electron-PhP interaction[4,5]:

$$g(\omega)=\frac{e}{\hbar\omega}\int_{-\infty}^{\infty}dz\,E_z(z)\,\mathrm{e}^{\left(-i\frac{\omega}{v}z\right)}$$

Figure 3B shows the results of fitting Eq. 1 with the measured α-$MoO_3$ PhP wavepacket and reference sample data, for 11 μm (top) and 12 μm (bottom) pump IR-pulses. These results reveal the wavelength-dependence of the PhP lifetime in this α-$MoO_3$ flake, which is 0.45ps for 11 μm (Figure 3B top) and shorter for 11.5 μm (not shown) and 12 μm (Figure 3B bottom). These results qualitatively agree with past experiments[32], showing inverse relation between lifetime and wavelength in this frequency range.

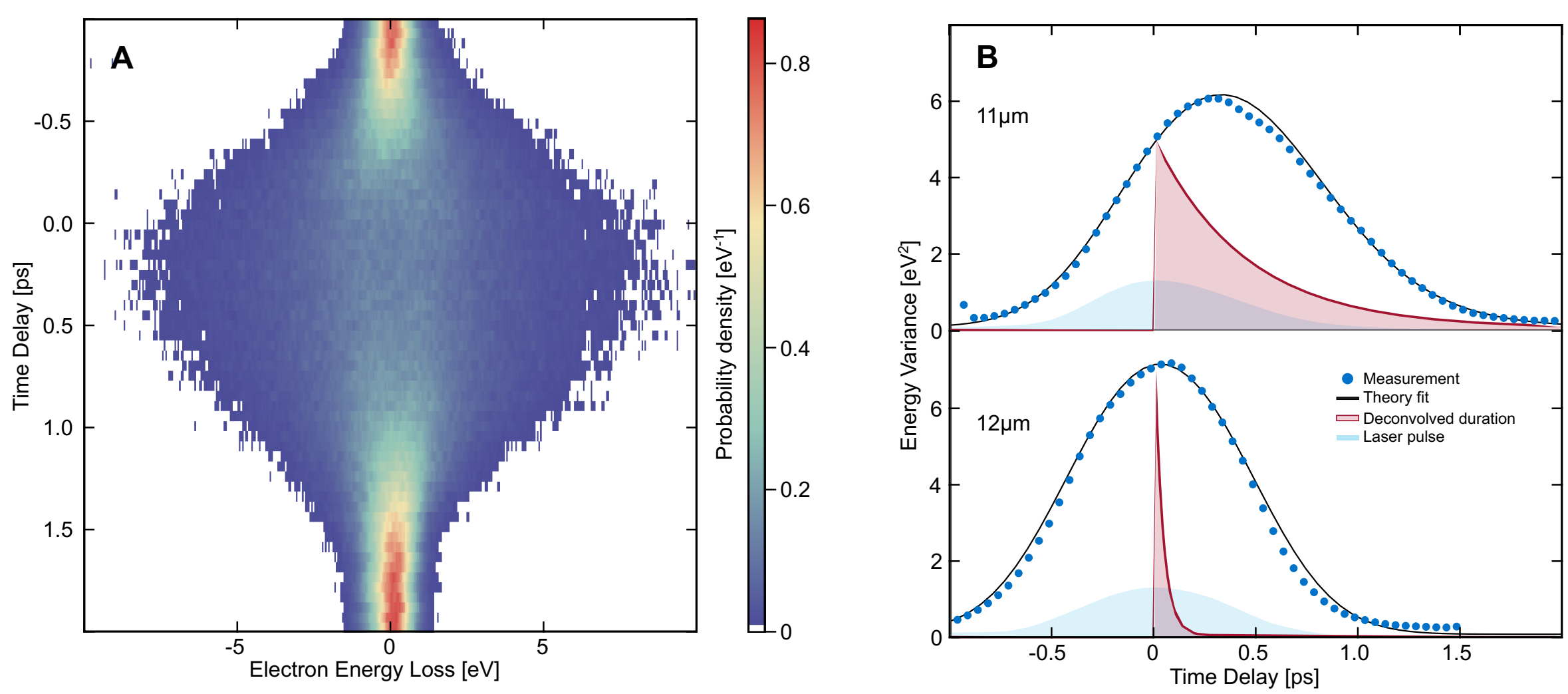

**Fig. 3. Measurement of the α-$MoO_3$ cavity lifetime via ultrafast electron energy spectroscopy. (A)** Electron energy spectrum as a function of time delay for an excitation pulse centered at 11 μm. **(B)** The electron energy variance as a function of time for 11 μm (top), and 12 μm (bottom). The extracted impulse response (red) shows the lifetime for each wavelength. The lifetime is strongly wavelength-dependent: $\tau$ = 0.45 ps for an 11 μm excitation pulse, versus $\tau$ = 0.04 ps (below our measurement resolution) for 12 μm. Blue shading: laser excitation profile measured using a reference sample.

## Discussion & Outlook

In summary, we extend the capabilities of PINEM further into the infrared and image the PhP wavepackets of α-MoO3 cavity-like flakes, revealing their spatiotemporal dynamics, and measuring wavelength-dependent lifetime. This experiment exemplifies the growing utilization of PINEM as a powerful tool for probing the intricate physics of 2D materials and their polaritons. Our study further underscores the advantages of PINEM, including its capacity to capture the spatiotemporal dynamics of polariton wavepackets and its overall sensitivity to electromagnetic fields, even those outside the sample boundaries.

Several key challenges remain to fully harness PINEM for 2D materials exploration. A primary obstacle is effectively coupling light into the materials. Unlike SNOM measurements where the probing tip is also used for effective light coupling, here the coupling is done from free-space, resulting in weak coupling to high k-vector components. This limitation prevents us from fully utilizing one of the key advantages of PINEM – its high spatial resolution. Conquering this challenge will necessitate innovative strategies, potentially involving nanopatterning techniques or insertion of another tip into the setup, to overcome the momentum mismatch.

Extending PINEM to longer wavelengths unlocks new possibilities for electron optics and for coherent manipulation of electron wavefunctions. Recent advancements have facilitated phase reconstruction and coherently amplified imaging through electron interferometry[16]. Fully exploiting the potential of electron interferometry necessitates a method to control both the temporal and spatial wavefunction of electrons. In that context, the polariton canalization and diffraction-less propagation observed in $MoO_3$ tri-layers[29], characterized by spectral robustness and all-angle tunability, may offer promising prospects for fabricating 2D-material devices capable of achieving this goal.

Looking forward, PINEM's high spatial resolution and the non-invasive nature of free-electron measurements makes this technique promising for measurements of highly confined polaritons with high momenta features[33,34]. Such polaritonic phenomena are elusive to measure by other means (e.g., in tip-based techniques, such as SNOM, the tip itself can be larger than the mode's wavelength and its proximity to the sample can be hard to control). Modes with such strong confinement are of particular interest for pico-cavities[35], where PINEM could enable measurements of unprecedentedly small cavities, provided a mechanism to couple such high momentum modes into the cavity. This capability would allow exploration of regimes where macroscopic material parameters such as permittivity may fail, and novel physical phenomena would emerge.

**Acknowledgements**

This project is funded by the European Union- ERC COG, QinPINEM, Project Number 101125662, by the Gordon and Betty Moore Foundation, through Grant GBMF11473, and by the Israel Science Foundation, Grant No. 385/23.